\documentclass[%
 reprint,
 amsmath,amssymb,
 aps,
]{revtex4-1}

\usepackage{graphicx}
\usepackage{dcolumn}
\usepackage{bm}

\setcitestyle{round}

\usepackage{siunitx}
\DeclareSIUnit{\molar}{M}
\usepackage[draft]{todonotes}   

\usepackage{color,soul}   

\usepackage{amsmath}
\usepackage{xcolor}

\begin{document}

\preprint{APS/123-QED}

\title{The effect of base pair mismatch on DNA strand displacement}

\author{D. W. Bo Broadwater Jr.}
\author{Harold D. Kim}%
\email{harold.kim@physics.gatech.edu}
\affiliation{%
 School of Physics, Georgia Institute of Technology\\
 837 State Street, Atlanta, GA 30332-0430
}%



\begin{abstract}
DNA strand displacement is a key reaction in DNA homologous recombination and DNA mismatch repair and is also heavily utilized in DNA-based computation and locomotion. Despite its ubiquity in science and engineering, sequence-dependent effects of displacement kinetics have not been extensively characterized. Here, we measured toehold-mediated strand displacement kinetics using single-molecule fluorescence in the presence of a single base pair mismatch. The apparent displacement rate varied significantly when the mismatch was introduced in the invading DNA strand. The rate generally decreased as the mismatch in the invader was encountered earlier in displacement. Our data indicate that a single base pair mismatch in the invader stalls branch migration, and displacement occurs via direct dissociation of the destabilized incumbent strand from the substrate strand. We combined both branch migration and direct dissociation into a model, which we term, the concurrent displacement model, and used the first passage time approach to quantitatively explain the salient features of the observed relationship. We also introduce the concept of splitting probabilities to justify that the concurrent model can be simplified into a three-step sequential model in the presence of an invader mismatch. We expect our model to become a powerful tool to design DNA-based reaction schemes with broad functionality. 
\end{abstract}

\maketitle

\onecolumngrid
\section{Introduction}

DNA strand displacement is a reaction where one of the strands in a double-stranded DNA is replaced with another nearly identical strand. It is a fundamental mechanism to exchange genetic material and plays an essential role in homologous recombination\cite{holliday1964mechanism} and mismatch repair\cite{jackson2002sensing,chapman2012playing}. DNA strand displacement involves three single strands named the invader, the incumbent, and the substrate strands and can be abstracted to a swapping reaction between the invader and the incumbent strands on the substrate strand. The invader can then be viewed as an input signal while the incumbent can be seen as an output signal. At this level of abstraction, DNA strand displacement can be idealized into ``tinker toys'' that fit together to form complex, interactive networks in the field of nanotechnology\cite{li2002new,green2014toehold,jiang2015robust,schulman2015designing} with applications in diverse areas such as biosensing\cite{zhang2012optimizing, zhang2014dna}, DNA construction\cite{marras2015programmable,zhou2015direct,rogers2014programming}, DNA motors\cite{liber2015bipedal,pan2015recent,landon2014energetically,tomov2013rational,masoud2012studying}, and DNA computation\cite{li2015spiking,wu2015survey,liu2014toehold,he2014dna,wang2014simple,qian2011scaling}. 

\begin{figure}[th!]
\includegraphics[width=8.5cm]{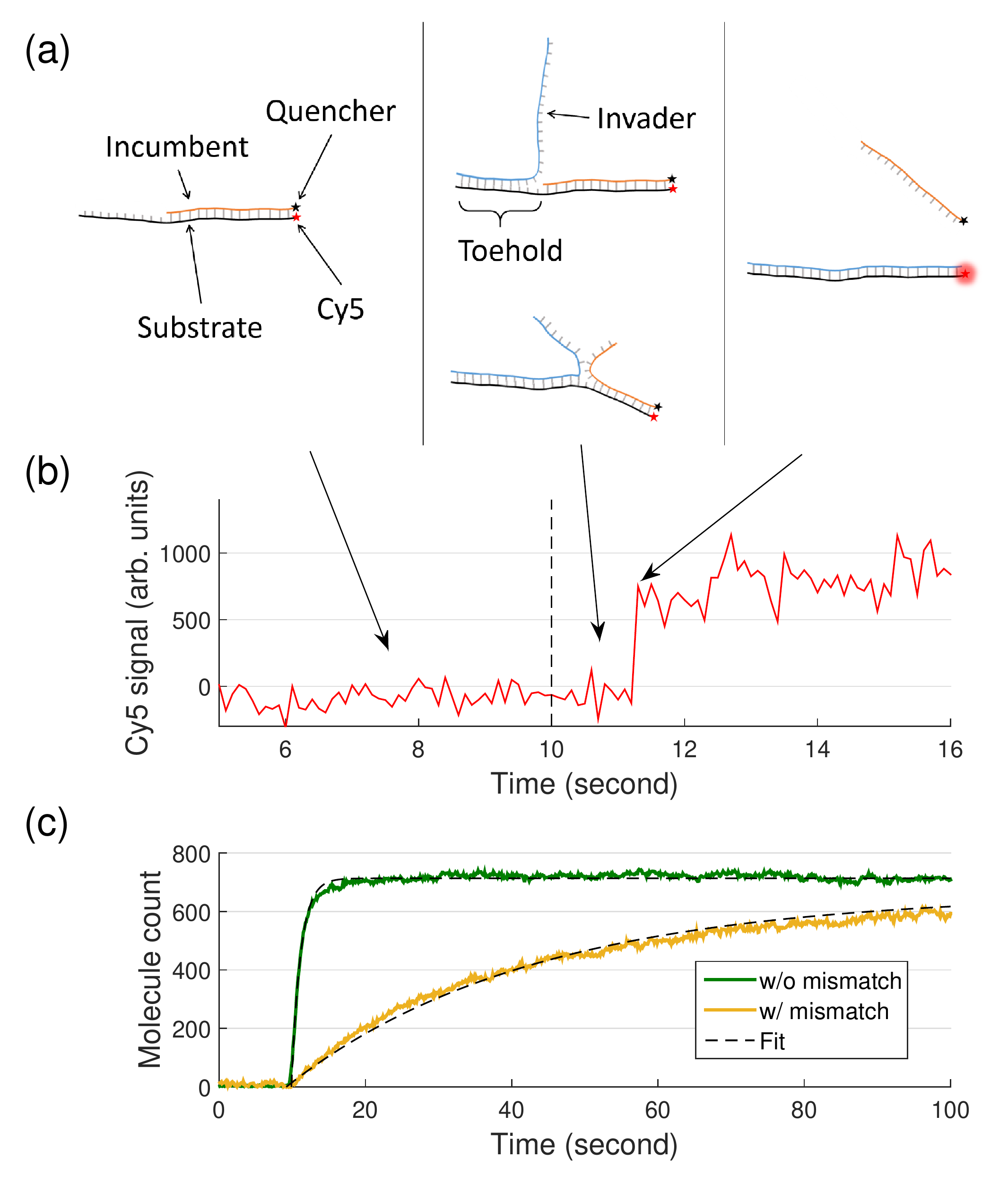}
\caption{Measuring strand displacement. (a) Experimental design. Three single strands of DNA termed substrate (black), incumbent (orange), and invader (blue) strands participate in strand displacement. Cy5 attached to the substrate is initially quenched due to the Black Hole Quencher on the incumbent. When the incumbent is displaced by the invader, Cy5 recovers its fluorescence. (b) Cy5 signal during strand displacement. Shown is the fluorescence time trajectory of a single Cy5 molecule obtained by total internal reflection microscopy. Invader molecules were introduced via flow (dashed line at the 10th second). A large, single, and sudden increase in fluorescence indicates displacement. (c) Extracting the apparent strand displacement rate. Two sets of sample data and their respective fits are plotted. Molecule count is calculated via in-house code scripted in MATLAB (The MathWorks, Natick, MA). The data are fitted to single exponential curves with an origin at the injection time (10 seconds after starting acquisition). Mean first passage times (MFPTs) are approximated as the reciprocal of the fitted rate constants.}
\label{fig:schematic}
\end{figure}

One class of strand displacements known as toehold-mediated DNA strand displacement is particularly useful because of sequence-dependent controllability. In this reaction, the shorter incumbent forms a partial duplex with the longer, complementary substrate (Fig.~\ref{fig:schematic}). The invader then hybridizes with the toehold, the unbound region of the partially-duplexed complement. The reaction is thought to proceed through a branch migration process until the incumbent is completely displaced\cite{srinivas2013biophysics}. The thermodynamics of this reaction is straightforward: the final state forms more canonical Watson-Crick base pairs and, therefore, must be lower in free energy than the initial state. In comparison, kinetics of strand displacement can vary by several orders of magnitude as a function of toehold length\cite{zhang2009control} and mismatch position\cite{machinek2014programmable}. 

However, current models of DNA strand displacement are either too simplified\cite{zhang2009control} or too detailed\cite{srinivas2013biophysics,machinek2014programmable} to capture position-dependent sequence effects on strand displacement kinetics. This study seeks to build a reaction scheme for toehold-mediated DNA strand displacement kinetics at the single base pair level. To construct this model, we measured the strand displacement rate in the presence of a mismatched base pair in the invader and the incumbent using single-molecule fluorescence. We found that a mismatch in the invader could dramatically slow down the strand displacement rate when positioned near the toehold. Based on this observation, we devised a reaction scheme that includes both branch migration and direct dissociation of the incumbent, which can be analyzed with ease using the first passage time approach. The observed dependence of strand displacement rate on mismatch position suggests that a single mismatched nucleotide in the invader can stall branch migration, and direct dissociation of the incumbent, but not complete branch migration, terminates DNA strand displacement. Our model analysis thus reveals direct dissociation of the incumbent as an essential pathway of DNA strand displacement. 

\section{Materials and Methods}
\subsection{Sample preparation}

Custom DNA oligomers were purchased from Integrated DNA Technologies (Coralville, IA), which were internally labelled near the $5'$ end with a Cy5 fluorophore to increase photostability\cite{lee2014internally} and a biotin linker at the $3'$ end for surface immobilization. The 26-nt sequence was chosen as a complement to a region of mRNA encoding Yellow Fluorescent Protein (YFP). The 14-nt incumbent sequences labeled with a BHQ-3 dark quencher at the 3' end were commercially synthesized by Biosearch Technologies (Petaluma, CA). The 24-nt invader sequences were purchased from Eurofins Scientific (Huntsville, AL). Single mismatch strands were chosen to preserve pyrimidine:purine ratio by exchanging $ G \leftrightarrow A$ and $T \leftrightarrow C$. The specific sequences are in Tables~\ref{table:immob-strands},~\ref{table:inc-strands}, and ~\ref{table:inv-strands} in the Supporting Material.

\subsection{Experimental setup}

Objective-type total internal reflection fluorescence microscopy (TIRFM) was implemented to image individual molecules. A commercially available microscope (IX81, Olympus, Melville, NY) was used to image Cy5 fluorophores excited by a 640 nm laser (CUBE 640- 30FP, Coherent, Santa Clara, CA). Binned images ($2\times2$) were captured with an EMCCD (DU-897ECS0-\#BV, Andor), and images were recorded at 10 fps with \SI{100}{\milli\second} exposure time using our in-house software. Experiments were performed on flow cells constructed as previously described\cite{le2014studying}, and a syringe pump (NE-1000, New Era Pump Systems) was used to control flow volume and flow rate (\SI{10}{\micro\liter/\second}).


The surface was passivated with biotinylated bovine serum albumin to minimize nonspecific binding. After Neutravidin coating, the Cy5-labeled substrate molecules were immobilized at \SI{50}{\pico\molar} in solution. Next, \SI{20}{\micro\liter} of dark quencher-labeled incumbent strands were pumped into the flow cell at \SI{200}{\nano\molar}. After \SI{5}{minutes}, excess dark quencher probes were washed away with oxygen scavenging imaging buffer\cite{aitken2008oxygen}, which contained \SI{1}{m\molar} 6-hydroxy-2,5,7,8-tetramethylchroman-2-carboxylic acid (Trolox), \SI{5}{m\molar} protocatechuic acid, \SI{100}{\nano\molar} protocatechuate 3,4-dioxygenase, \SI{100}{m\molar} Tris-HCl (pH 7), and \SI{300}{m\molar} NaCl. Strand displacement was initiated by pumping invader strands in imaging buffer at \SI{2}{\micro\molar} into the flow cell. 

As the incumbent was displaced, fluorescent signal reappeared. The reappearance of fluorescent signal was recorded and analyzed using in-house MATLAB software. The molecule count (cumulative sum) was fitted to a single exponential curve and from that curve an overall rate of strand displacement was extracted. The mean first passage time was estimated as the reciprocal of the extracted rate. The experiment was repeated in triplicate for all single mismatch strands derived from the perfectly matched incumbent and invader.

\subsection{Concurrent displacement model}

We define a 1D lattice with $n$ sites, where $n$ is the number of bases in the incumbent strand. For simplicity, we assume that the rate of breaking individual base pairs is slower than the reverse rate of formation. Under this assumption, the incumbent and the invader would remain completely zippered up with the substrate strand. Therefore, we can specify each intermediate state with one state variable $i$, which is equal to the number of displaced base pairs. For example, $i=0$ represents the state where the invader has not displaced any base pair, and $i=n-1$ corresponds to a state where the invader has displaced all but one base pair between the incumbent and the substrate. We add two boundary states (N and V) to this Markov chain. N stands for the in`c'umbent only state, and V for the in`v'ader only state. Branch migration at i-th lattice site is performed in single steps at forward and reverse rates, $f_i$ and $r_i$ respectively. It is important to note that these rates are expected to be much slower than the single base pair opening (fraying) rate because a single fraying event does not necessarily lead to branch migration. As it stands, this model is equivalent to a random walk with a perfectly reflecting boundary on the left (N) and perfectly absorbing boundary on the right (V). Mean first passage time (MFPT) of this 1D model can be easily derived with or without a kinetic barrier (See Supporting Material).

\begin{figure}[th!]
\includegraphics[width=8.5cm]{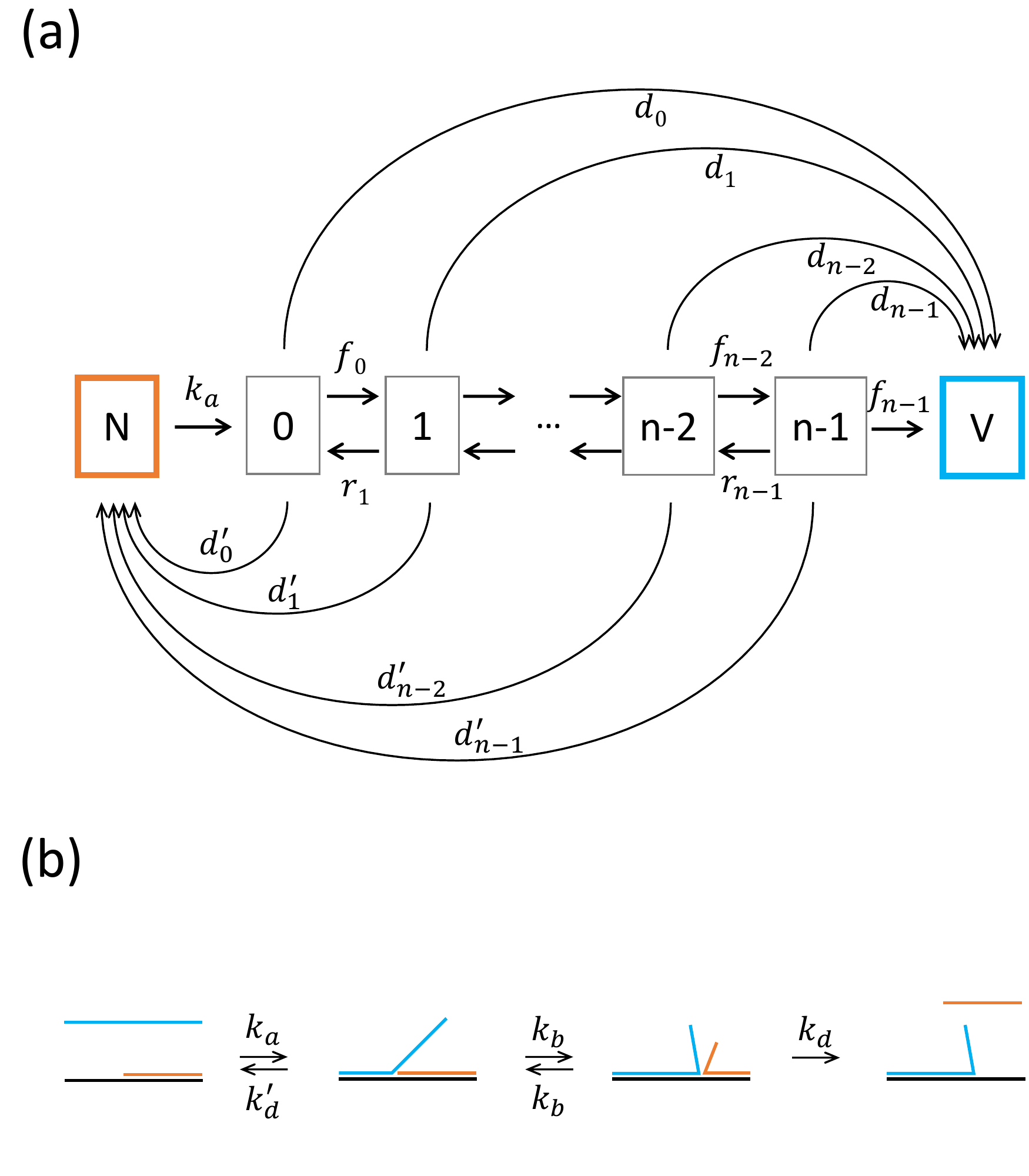}
\caption{Strand displacement models. (a) Concurrent displacement model (``yin-yang'' model). Strand displacement is preceded by a state occupied by the incumbent only (N). Branch migration begins after the toehold annealing step ($k_a$). The branch point can take any value ($i$) between $0$ and $n-1$ and migrate towards nearest neighbors with forward rate ($f_i$) and reverse rate ($r_i$). V is the invader only state. Concurrently with branch migration, the invader and the incumbent can dissociate from any intermediate state with rates ($d'_i$ and $d_i$) that depend on the number of base pairs . (b) Sequential displacement model. The initial state and the final state are identical to N and V. The orange, black, and blue line segments represent the incumbent, the substrate, and the invader, respectively. The invader may anneal ($k_a$) and dissociate ($k_d'$) with the toehold reversibly. Afterwards, branch migration ($k_b$) proceeds until a mismatch is encountered at state $j-1$ or the incumbent strand is significantly destabilized at state $n_{th}$. From either state, the incumbent can irreversibly dissociate ($k_d$).} 
\label{fig:model}
\end{figure}

We can extend this model further to include direct dissociation of the invader and incumbent. As shown in Fig.~\ref{fig:model}(b), the invader and the incumbent can dissociate at rates $d'_i$ and $d_i$, respectively, from each state. This model scheme is thus similar to a general kinetic proofreading scheme\cite{bel2010simplicity} with additional feedforward paths to the final state. Given the toehold length of $n_t$, the invader is held by $n_t+i$ base pairs in state $i$ and, therefore experiences a decrease in dissociation rate as more base pairs are formed. Conversely, since the incumbent is held by $n-i$ base pairs in state $i$, the dissociation rate would become larger as more base pairs are displaced. According to the previous work by Anshelevich et al.\cite{anshelevich1984slow}, the relationship between duplex dissociation rate ($k_d(N_{bp})$) and the number of base pairs ($N_{bp}$) is given by
\begin{equation}
k_d(N_{bp})=\frac{2k_0N_{bp}}{s^{N_{bp}-1}},
\label{eq:kamen}
\end{equation}
where $s$ is termed the stability factor equal to the ratio of rates of closing to opening for a single base pair, and $k_0$ is the unzipping rate of a single base pair at the melting temperature. This expression is essentially identical to the expression used by Zhang and Winfree\cite{zhang2009control}. 

The MFPT of the concurrent model requires consideration of the master equation
\begin{equation}
\frac{d \mathbf{x}}{dt}=-\mathbf{A}\mathbf{x} ,
\label{eq:master}
\end{equation}
where $\mathbf{x}$ is defined as the $(n+1) \times 1$ state vector with probabilities in each state ($x_j(t)$) as components, and $\mathbf{A}$ is an $(n+1) \times (n+1)$ transition matrix, which is nearly tridiagonal, with components:
\begin{equation}
\mathbf{A} = \begin{bmatrix}
k_a+d_0 &      -d'_0      &        -d'_1     &       -d'_2      & \cdots &           -d'_{n-2}              &         -d'_{n-1}        \\
-k_a    &  f_0+d_0+d'_0   &        -r_1      &         0        & \cdots &                0                 &             0            \\
  0     &       -f_0      & f_1+r_1+d_1+d'_1 &       -r_2       &  \cdots      &             0               &          0          \\
  0     &         0       &         -f_1     & f_2+r_2+d_2+d'_2 & \cdots &                0                 &             0            \\
  \vdots     &         \vdots       &         \vdots       &       \vdots       & \ddots &             \vdots             &             \vdots           \\
0  &     0      &      0      &      0            & \cdots  & f_{n-2}+r_{n-2}+d_{n-2}+d'_{n-2} &        -r_{n-1}          \\
  0     &         0       &          0       &      0      &    \cdots  &             -f_{n-2}             & r_{n-1}+d_{n-1}+d'_{n-1} \\
\end{bmatrix} .
\end{equation}
If the initial condition is given by $x_j(0)=\delta_{j1}$, the MFPT ($\tau$) can be expressed with matrix determinants as\cite{kim1958mean}
\begin{equation}
\tau=\frac{
\begin{vmatrix}
1 & 1 & \cdots &1\\
a_{21} & a_{22} & \cdots &a_{2n}\\
\vdots & \vdots & \ &\vdots\\
a_{n1} & a_{n2} & \cdots &a_{nn}
\end{vmatrix}}{
\begin{vmatrix}
a_{11} & a_{12} & \cdots &a_{1n}\\
a_{21} & a_{22} & \cdots &a_{2n}\\
\vdots & \vdots & \ &\vdots\\
a_{n1} & a_{n2} & \cdots &a_{nn}
\end{vmatrix}} ,
\label{eq:3dmodel}
\end{equation}
where $a_{ij}$ is the element of $\mathbf{A}$. 

Without a mismatch, branch migration rate is the same in both directions ($f_{i-1}=r_i$). In comparison, a mismatch in the incumbent speeds up forward migration by a ratio $a$, and a mismatch in the invader speeds up reverse migration by the same ratio. $a$, termed the mismatch migration ratio should be much larger than one. Assuming that branch migration rates are identical to $f$ for all nucleotides without a mismatch, the MFPT can be uniquely determined with five parameters $k_a$, $k_0$, $s$, $f$, and $a$. 


\subsection{Splitting probabilities}
For the concurrent model, it is of particular interest to ask from which state the incumbent dissociates most frequently. This concept is related to splitting probabilities in stochastic processes. In our model, the incumbent strand can reach the absorbing boundary state (V) from $n+1$ different states. The time dependence of V is given by the rate equation 
\begin{equation}
\frac{dV(t)}{dt}=\sum_{j=1}^{n+1}k_jx_j(t)
\end{equation}
where $k_j$ is the dissociation rate constant from the j-th state. In the long time limit, the system will be completely depleted of the incumbent through $n+1$ channels. 
\begin{equation}
V(t\rightarrow\infty)=\sum_{j=1}^{n+1}\int_0^\infty k_jx_j(t)dt\equiv\sum_{j=1}^{n+1} \pi_j, 
\end{equation}
where $\pi_j$ is the splitting probability through the j-th state. Using eigen-decomposition of $\mathbf{A}$, it is straightforward to show
\begin{equation}
\pi_j=k_j\int_0^\infty x_j(t)dt=k_j\int_0^\infty \sum_{i=1}^{n+1}(e^{-\mathbf{A}t})_{ji}x_i(0)=k_j\sum_{i=1}^{n+1}(\mathbf{A}^{-1})_{ji}x_i(0), 
\end{equation}
where $\mathbf{A}^{-1}$ is the inverse matrix of $\mathbf{A}$. Thus, with the initial condition $x_j(0)=\delta_{j1}$, we obtain the following formula for splitting probabilities
\begin{equation}
\pi_j=k_j \frac{\mathbf{C}_{1j}}{|\mathbf{A}|},
\label{eq:split}
\end{equation}
where $\mathbf{C}$ is the cofactor matrix of $\mathbf{A}$. $\pi_j$ is also related to the mean first passage time (Eq.~\ref{eq:3dmodel}) according to
\begin{equation}
\tau=\sum_{j=1}^{n+1}\frac{\pi_j}{k_j}.
\end{equation}

\subsection{Sequential displacement model}
One can also build a three-step sequential model (Fig.~\ref{fig:model}(b)) that qualitatively captures the effect of a mismatch on strand displacement. The first step is toehold formation through annealing ($k_a$) accompanied by reverse dissociation ($k'_d$). The second step is reversible branch migration ($k_b$). The third step is dissociation of the incumbent ($k_d$), which is irreversible in our experiment. The key difference of the model from the concurrent model is that branch migration and dissociation occur in a serial fashion. The mean first passage time ($\tau$) for this reaction is given by 
\begin{equation}
\tau\approx\frac{1}{k_a}+\frac{1}{k_b}+\frac{2}{k_d}. 
\label{eq:fullthreestep}
\end{equation}
This equation can be derived from either Eq.~\ref{eq:3dmodel} or Eq.~\ref{seq:1dmfpt} under the approximation that the invader association rate is faster than the dissociation rate ($k_a \gg k_d^\prime$). The third step (incumbent dissociation) can occur from a state where (i) branch migration is stalled due to a mismatch in the invader, or (ii) the incumbent-substrate interaction is significantly weakened, with only a few intact base pairs left between them. We model that $n_{th}$ number of base pairs have to be displaced for the incumbent to dissociate. The branch migration rate ($k_b$) also depends on migration distance ($N_{bp}$), which can be derived from the mean first passage time of a standard one-step process (Eq.~\ref{eq:1dmatch}) as
\begin{equation}
k_b(N_{bp})=\frac{2f}{N_{bp}(N_{bp}+1)}.
\end{equation}
If the mismatch in the $j$-th position is encountered before $n_{th}$, branch migration stalls at state $j-1$. In the absence of a mismatch, branch migration continues till the threshold state $n_{th}$. Whichever occurs first becomes the state where the incumbent dissociates ($\min(j-1,n_{th})$). Therefore, we can express the dependence of Eq.~\ref{eq:fullthreestep} on mismatch position as 
\begin{equation}
\tau=\frac{1}{k_a}+\frac{1}{k_b(\min(j-1,n_{th}))}+\frac{2}{k_d(n-\min(j-1,n_{th}))}.
\label{eq:threestep}
\end{equation}

\subsection{Data fitting}
Nonlinear least squares fitting was performed with `lsqcurvefit' of the MATLAB Optimization Toolbox. Eq.~\ref{eq:3dmodel} was used as the fitting function. All individual measurements were fitted with equal weight using shared fitting parameters. These measurements include mismatch in the invader and the incumbent as well as the perfect match strand.   

\section{Results}

\begin{figure*}[th!]
\includegraphics[width=17cm]{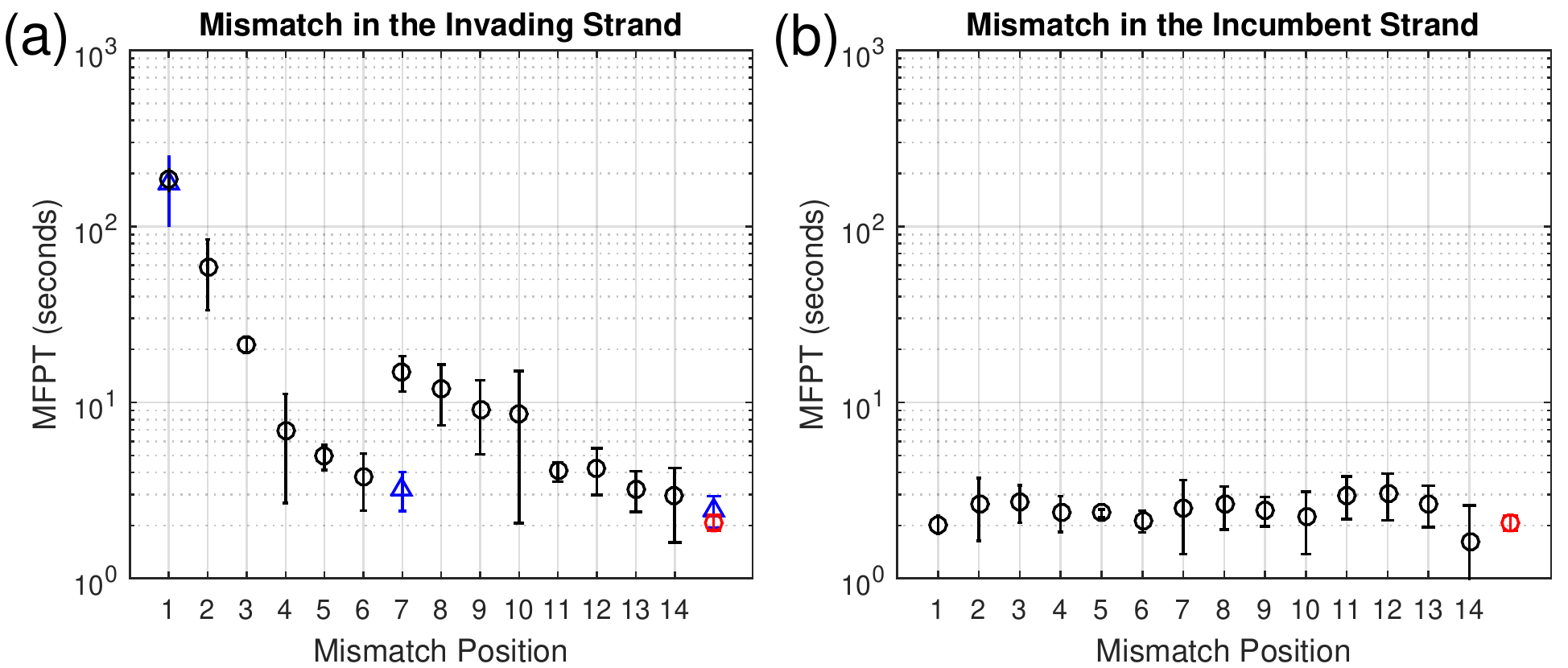}
\caption{Observed displacement MFPTs. Molecule number was counted as a function of time and fitted to a single rate, and MFPT was calculated as the reciprocal of that rate. We plot the average and standard deviation of observed MFPT against the mismatch position in both the invader and incumbent. The perfect match case is plotted in red. The strands designed to be free of secondary structure are plotted in blue. (a) Invader mismatch. The MFPT generally increases with mismatch position and shows a nearly 50-fold variation overall. Notably, at positions 7, 8, and 9, the MFPTs are higher than the overall trend. The MFPTs for strands without secondary structure align closely with their counterparts for the match strand and mismatch position 1 strand, but there is a clear distinction for the mismatch position 7 strands. (b) Incumbent mismatch. In comparison to (a) the mismatch position does not have a significant effect on the MFPT.}
\label{fig:data}
\end{figure*}

We performed toehold-mediated strand displacement by challenging a surface-immobilized substrate-incumbent partial duplex with invader strands free in solution. In this experimental scheme, every reaction step can be treated as first order. Formation of partial duplexes between the substrate and the incumbent on the surface led to disappearance of most Cy5 spots due to quenching. Upon perfusion of the invader, Cy5 spots reappeared over time, which was interpreted as strand displacement (supplemental movie S1). We counted individual spots over time and extracted the apparent displacement rate from single exponential fitting. We performed this experiment in triplicate for 15 invaders (14 mismatch strands + match strand). In Fig.~\ref{fig:data}(a), we plotted the strand displacement rates measured for each mismatch position. It took \SI{\sim2}{\second} for a perfectly matching invader to displace the incumbent (red point, Fig.~\ref{fig:data}(a)). When the mismatch was introduced in the invader, strand displacement became slower, especially for the first four positions near the toehold region. The relationship was overall monotonic (except strands 7 through 10) with a roughly 70-fold change in the observed rate between the strands with a mismatch in the first and last positions. The effect of the invader mismatch is the strongest at the first position, but seems to be significantly weakened by position 6.  

We suspected that the deviation from this trend at positions 7 through 10 might stem from a secondary structure in the invader. Strand 7 and 8, for example, are predicted to form stable hairpins (Supplementary Fig.~\ref{sfig:secondary-struct}). Thus, we designed new sequences free of secondary structure for another set of strand displacement experiments. The invaders we tried were perfect match and mismatches at position 1 and 7. The rate measured with the new mismatch 7 strand was significantly faster, similar to the baseline, whereas the rates measured with the new perfect match and mismatch 1 strands remained unchanged. This result lends support to our speculation that the deviant points are caused by secondary structure.

We performed a similar experiment to explore the relationship between displacement rate and single mismatch position on the incumbent strand. In Fig.~\ref{fig:data}(b), we plotted the measured strand displacement rates for each mismatch position. In contrast to the dynamic pattern for the invader mismatch, there is relatively little variation in displacement rate against the mismatch position in the incumbent strand. All rates were similar to the displacement rate without a mismatch (red point, Fig.~\ref{fig:data}(b)). This incumbent mismatch experiment serves to control for the possibility of interacting dangling ends since the same dangling ends are available to interact in both the invader and incumbent mismatch. The lack of variation in rate over mismatch position for the incumbent implies that the observed complex behavior for the invader mismatch (Fig.~\ref{fig:data}(a)) is not due to interacting dangling ends.

These two sets of experiments corroborate the intuition that it is easier to displace a mismatch strand with a match strand than vice versa. Also, it confirms the previous inference of branch migration rate of \SI{1}{\per\second} over a similar length of displacement domain\cite{zhang2009control}. In contrast, a mismatch in the invader can dramatically slow down strand displacement, especially, if placed near the toehold. This implies that the mismatch effect could be modeled as a localized kinetic barrier that disrupts zipping of the invader. 

To understand the mismatch effect in a quantitative fashion, we attempted to model toehold-mediated strand displacement as a one-dimensional random walk\cite{murthy1989mean,pury2003mean} with a single misstep (See Supporting Material). This model assumes that displacement of the incumbent occurs via complete branch migration towards the boundary. The branch point moves much faster forward upon an incumbent mismatch and much faster backward with an invader mismatch. Derivation of the mean first passage times (MFPT) is straightforward for both invader and incumbent mismatches, and the analytical formula are presented as Eq.~\ref{eq:1dmodel} in the Supporting Material. However, this model incorrectly predicts a parabolic dependence of rate on mismatch position, with the mismatch in the center having the most significant effect\cite{lakatos2004first}. While further inclusion of a slow initiation step causes the model to produce a general monotonic trend (Fig.~\ref{sfig:1dmodel} in the Supporting Material), it still cannot produce the sharp drop in rate observed over the first few positions.

The failure of the one-dimensional model prompted us to extend the model by including direct dissociation of the incumbent even before the branch point reaches the end. We reasoned that these direct dissociation paths would become important especially when branch migration is hindered or when the incumbent binding is severely weakened near the latter stage of branch migration. This model is termed the concurrent displacement model, and is schematized in an aesthetic yin-yang pattern as shown in Fig.~\ref{fig:model}(a).

\begin{figure*}[th!]
\includegraphics[width=17cm]{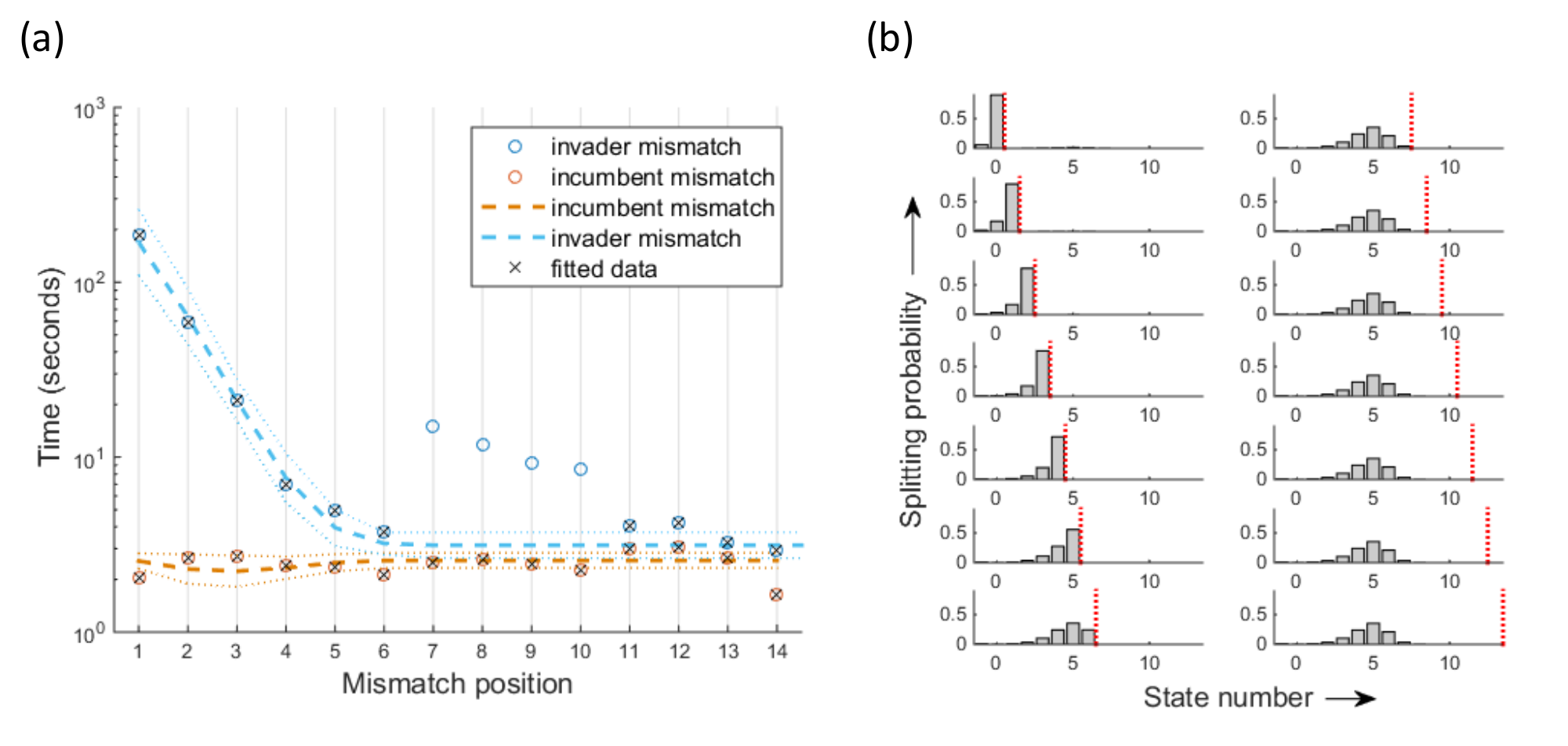}
\caption{Model analysis. (a) Fit for concurrent displacement model. The plot shows the relationship between the apparent mean displacement time vs. the mismatch position in the invader (blue) and the incumbent (orange). We used five fitting parameters, the annealing rate ($k_a$), the dissociation rate constant ($k_0$), the branch migration rate ($f$), the stability factor ($s$), and the mismatch migration ratio ($a$). The points included in the fitting routine are marked by `$\times$'. The dashed lines show the fit by our model, and the dotted lines are \SI{95}{\percent} confidence intervals at the input values. (b) Splitting probability distribution. The bar graphs show the splitting probability vs. state number with the mismatch position varying from 1 to 7 (top to bottom, left), and from 8 to 14 (top to bottom, right). The parameters obtained from the fit in (a) were used to calculate the probabilities. The red vertical dash indicates the mismatch position.}   
\label{fig:modelfit}
\end{figure*}

We performed nonlinear least squares fitting of MFPT (Eq.~\ref{eq:3dmodel}) to the measured displacement times. The dissociation rate ($d_0'$) of the invader from the toehold was directly measured (Supplementary Fig.~\ref{sfig:binding}) and constrained in the fitting. All other invader dissociation rates ($d_{i=1,2,...,n-1}'$) were expressed in relation to the measured value $d_0'$ by a single parameter $s$ according to Eq.~\ref{eq:kamen}. The 4 data points at mismatch positions 7, 8, 9, and 10  that markedly deviated from the monotonic pattern were omitted, which we justify based on our additional measurement with strands rationally designed to be free of secondary structures (blue triangles, Fig.~\ref{fig:data}). As shown in Fig.~\ref{fig:modelfit}(a), the concurrent displacement model can well fit both observed relationships with a common set of parameters. The mismatch migration ratio ($a$) diverges, and therefore an upper bound was placed. The best fit produces the association rate ($k_a$) of \SI{0.6}{\per \second}, the dissociation rate constant ($k_0$) of \SI{3e5}{\per \second}, the branch migration rate ($f$) of \SI{10}{\per\second}, and the stability factor ($s$) of $5.1$.    

The association rate, \SI{\sim0.6}{\per\second} at \SI{\sim2}{\micro\molar}, is similar to the association rate constant measured in bulk (\SI{\sim1}{\per\micro\molar\per\second}) considering the surface effect\cite{gao2006secondary} or differences in salt condition or temperature\cite{zhang2009control,cisse2012rule}. This value is also close to the association rate (\SI{1}{\per\second} at \SI{2}{\micro\molar}) inferred from our separate measurement of concentration dependence (Supplementary Fig.~\ref{sfig:binding}). The extracted branch migration step time is \SI{100}{\milli\second}. This is seemingly much longer than \SI{2.5}{\milli\second} previously inferred based on the three-step displacement model\cite{zhang2009control}. This disparity, however, is not due to different measurements of apparent branch migration rates ($k_b$), but likely due to different models used to infer the step rate ($f$). For example, the apparent time it takes to displace a 14-nt domain in our experiment is \SI{\sim1}{\second} (red point in Fig.~\ref{fig:data})), similar to the inferred branch migration rate of \SI{\sim1}{\second} over a 20-nt domain\cite{zhang2009control}. The spontaneous unzipping rate of a single base pair ($k_0$) is estimated to be \SI{e6}{\per\second} to \SI{e7}{\per\second}\cite{anshelevich1984slow,srinivas2013biophysics}. Our estimate of $3\times10^5 s^{-1}$ is within an order of magnitude, and is also similar to a thermodynamic estimate (\SI{6e5}{s^{-1}}) used by Zhang and Winfree\cite{zhang2009control}. Finally, the extracted stability factor ($s$) is 5.1, which indicates that the base pair is 5.1 times more likely to close than open. This ratio is close to $10^{0.6}$ obtained by extrapolation of a semi-analytical calculation\cite{cocco2001force}. 

To gain more insights into the mismatch effect, we calculated the probability that the incumbent strand dissociates from each state using the parameters obtained from fitting to the concurrent model. This probability is conceptually similar to the splitting probabilities between different absorbing states in a one step process\cite{van1992stochastic}. In our concurrent model, the splitting probabilities leading to the absorbing state V can be calculated using Eq.\ref{eq:split}. In Fig.~\ref{fig:modelfit}(b), the splitting probability for each state $i$ is plotted as a bar graph with varying mismatch positions marked by red vertical lines. As expected, the splitting probabilities sum to one in all cases. For early mismatch positions (left half, Fig.~\ref{fig:modelfit}(b)), splitting probabilities past the mismatch position are zero, which indicates that branch migration does not proceed beyond the mismatch. For late mismatch positions (right half, Fig.~\ref{fig:modelfit}(b)), the incumbent dissociation is complete even before the mismatch is encountered, which explains why the displacement rate is not affected by the mismatch. The key insight from this model analysis is that the invader mismatch stops branch migration, and displacement is terminated by incumbent dissociation, not by branch migration.    Based on this insight, we can build a simpler sequential displacement model (Fig.~\ref{fig:model}(b)) to rationalize the observed dependence of strand displacement rate on mismatch position. The MFPT of this reaction scheme is expressed as a sum of three terms, association time, branch migration time, and dissociation time (Eq.~\ref{eq:threestep}). The position dependence mainly arises from the third term, which decreases with increasing mismatch position only up to some threshold state ($n_{th}$) and remains unchanged beyond it.

\section{Discussion}
In this study, we designed a surfaced-based single-molecule assay to measure kinetics of toehold-mediated DNA strand displacement and its dependence on a base pair mismatch. In contrast to bulk measurements\cite{zhang2009control,srinivas2013biophysics,machinek2014programmable}, our assay produces the strand displacement rate from a first-order reaction, which does not depend on substrate concentration. Furthermore, due to the long toehold and high invader concentration used in our assay, strand displacement is completed in a time scale of a few seconds (or minutes with a mismatch), significantly faster than typical bulk experiments. Due to this high efficiency, we expect our experimental method to become a powerful tool for the study of the biophysics of branch migration. Although not exploited in this study, the single-molecule aspect of our method can produce the full distribution of individual strand displacement times as well\cite{bel2010simplicity,floyd2010analysis}, which will be a topic of our future study.


Recently, the effect of an invader mismatch was studied in bulk\cite{machinek2014programmable}. This study found a similar qualitative dependence of displacement rate on mismatch position to ours, and hinted direct incumbent dissociation as an important pathway for displacement. The authors made an extra effort to preserve the trinucleotide sequence around a mismatch to minimize variation in mismatch free energy, which led to omission of some mismatch positions including the first. A qualitative explanation based on dynamics simulation was given, but a quantitative model predictive of displacement rates was missing. Our study thus complements the previous study by testing the mismatch effect in both the invader and the incumbent, at different positions, and with different DNA sequences. More importantly, we present a quantitative model and a first passage time approach to rationalize the mismatch effect.

Our concurrent model uses an intermediate level of coarse-graining compared to two types of previous models for toehold-mediated strand displacement. The first type is the three-step displacement model, which breaks the reaction into bimolecular toehold binding, unimolecular branch migration, and unimolecular dissociation from the final state\cite{zhang2009control}. The second type is a more detailed model at the molecular level, which includes intermediate states during branch migration\cite{srinivas2013biophysics}. Because the three-step model coarse-grains all of branch migration into a single step, it cannot readily incorporate nucleotide-level effects. The second model is thorough, but the implementation and analysis of this model require dynamics simulations with constraints and pre-equilibrium assumptions\cite{machinek2014programmable}. In comparison, our concurrent displacement model is straightforward to analyze using the first passage time approach presented here.

A complete understanding of the concurrent model requires solving the master equation (Eq.\ref{eq:master}). However, because the system has many intermediate states with transition rates of similar magnitudes, the time dependence of the probability distribution ($x_j(t)$) is expected to be characterized by multiple exponential terms, and cannot be easily subjected to fitting analysis. The first passage time (MFPT) approach bypasses this technical difficulty. Unlike $x_j(t)$, calculation of the MFPT can be easily done without solving the master equation. Moreover, splitting probabilities can be easily obtained as well. The quantitative framework we employed here can thus be applied to any other complex reaction scheme.

In fitting a five parameter model to 24 data points, we recognize that precise values of each parameter become difficult to determine. This limited range of data points is inherent to the nature of the experiment. A larger data set requires a longer invader, which would become more susceptible to secondary structure formation and more spurious intermolecular interactions. Further, sequence dependence in individual steps of branch migration can lead to deviations from our model prediction. Nevertheless, fitting parameters $k_0$ and $s$ are in agreement with other studies, and $k_a=\SI{0.6}{\per\second}$ is similar to our own estimate of $\SI{1}{\per\second}$ (Supplementary Fig.~\ref{sfig:binding}). Based on the extracted $k_0$ and $s$, the dissociation rate of the incumbent from state 4 is predicted to be \SI{2.6}{\per\second}. If complete base pairing of the substrate is assumed, state 4 corresponds to 10 base pairs between the incumbent and the substrate. But the dissociation rate of the 10-bp duplex between the invader and the substrate was measured to be much slower at \SI{1/30}{\per\second}. This comparison suggests that in state 4, the incumbent-substrate interaction is markedly destabilized probably due to a repulsive interaction between the incumbent and the invader near the branch point\cite{betterton2005opening}. Interestingly, Srinivas et al.\cite{srinivas2013biophysics} found that branch migration intermediates are destabilized by $3.4 k_BT$ due to dangling ends, which corresponds to $\sim30$-fold change in dissociation rate. Our model analysis is thus consistent with the incumbent having effectively 2-3 fewer intact base pairs than indicated by the location of the branch point. In other words, the incumbent is not completely zippered up against the invader.


Our own estimation of branch migration time of $\SI{\sim100}{\milli\second}$ per base pair step, notwithstanding measurement and fitting uncertainty, is much slower than predictions in the literature based on base pair fraying and one dimensional rate models\cite{zhang2009control,srinivas2013biophysics}. But, we think our estimation is plausible for a few reasons. A single fraying event is not likely to provide enough room or time for a base on another strand to invade. Thus, invasion of a base \textit{in trans} should occur only after many fraying/unfraying events, which could explain our slow branch migration rate. Also, 1D rate models are expected to significantly overestimate migration rates by omitting direct dissociation pathways which are the dominant mechanisms of displacement. To the best of our knowledge, branch migration step time has not been measured directly in the specific context for strand displacement. The branch migration time in Holliday junctions is estimated to be faster at $\SI{3.6}{m\second}$\cite{panyutin1994kinetics}, but it is not accompanied by a pair of free dangling ends that can destabilize the branch point due to crowding\cite{srinivas2013biophysics} or thermal fluctuation\cite{waters2015calculation}. Also, the low ionic strength used in our experiment may reduce branch migration rate as low salt would lead to slower base pair formation\cite{cisse2012rule}. A different experimental strategy that prevents or decouples direct dissociation of the incumbent is certainly necessary to accurately measure the branch migration rate.  

We used high concentrations of invader strand to speed up strand displacement and to minimize variability in our measurement. We measured the displacement rate for the perfect match invader as a function of concentration (Supplementary Fig.~\ref{sfig:binding}) and selected a concentration significantly above the midpoint (\SI{\sim2}{\micro\molar}). In this regime, the displacement rate is relatively insensitive to variation in concentration, and allows us to compare rates with different invaders. It was recently shown that DNA duplex can dissociate by competing complementary single strands without toehold mediation\cite{paramanathan2014general}, but this effect kicks in at a much higher concentration (\SI{\sim50}{\micro\molar}). Furthermore, the rates that we observed are at least an order of magnitude faster. Therefore, this mechanism cannot be relevant to our observations.

We have not comprehensively investigated the origin of deviation seen in invader mismatch strands 7,8,9, and 10. These strands exhibit significantly slower displacement rates than the rest, which led us to consider secondary structure formation. Using the secondary structure prediction program mfold\cite{zuker2003mfold}, we find that invader strands 2, 3, 7, and 8 adopt a relatively stable secondary structure (Fig.~\ref{sfig:secondary-struct} in the Supporting Material). Such secondary structure can negatively impact strand displacement at the annealing step. In addition to internal secondary structure, individual strands can form transient base pairs with one another, which could also retard the branch migration rate. The reduction in rate due to secondary structure would be more noticeable for 7 and 8 where strand displacement is fast. Furthermore, measurements for strands designed without secondary structure were much closer to the expectation (Fig.~\ref{fig:data} blue). We could not account for the origin of outliers at position 9 and 10 (Fig.~\ref{fig:data}).

Our method is not without limitations. First, we infer MFPT by exponential extrapolation from the appearance of fluorescence signal. In theory, MFPT cannot be measured if the initial population size is not known or if the reaction is not complete. Nonetheless, all displacement kinetics curves and their exponential extrapolations have similar integrated areas (See Fig.~\ref{fig:schematic}(c)), which indicates that our MFPT estimation is accurate. Second, Cy5 and the quencher in our experimental design might stack with each other or with neighboring bases to affect the intrinsic dissociation kinetics of the incumbent strand. However, such stabilizing interaction would only attenuate position-dependent mismatch effect, if any. Moreover, a similar experimental design did not affect the apparent displacement rate\cite{machinek2014programmable}. Therefore, the main conclusion we draw based on our model is likely valid.

{Our results have interesting implications for related areas in biology. Given the role of strand displacement in homologous recombination, it is conceivable that the repair mechanism could be affected by a single base mismatch in a position dependent manner. In a more applied sense, position dependence of strand displacement rate could be exploited to design masked probes for single-nucleotide variant detection in vitro or in vivo with increased specificity.

\section{Conclusion}
We used a novel experimental strategy to study toehold-mediated DNA strand displacement as a first-order reaction. At the single-molecule level, we measured the apparent displacement rate through recovery of fluorescence and found its strong dependence on mismatch position in the invader. We rationalized the observed dependence using the concurrent displacement model that allows branch migration and dissociation of the incumbent. Our model analysis suggests that a single base pair mismatch in the invader poses an almost insurmountable kinetic barrier to branch migration and reveals direct dissociation of the destabilized incumbent as the dominant pathway for displacement. We anticipate our kinetic model, which we colloquially term the yin-yang model, and the first passage time approach to be highly relevant to an understanding of dynamic response for an expansive range of complex networks.     

\section{Author Contributions}
D.W.B.B. and H.D.K. designed the research, analyzed the data, and wrote the paper. D.W.B.B. performed experiments.

\section{Supporting Citations}
Reference \cite{mentes2015free} appears in the Supporting Material.

\section{Acknowledgements}
This work was supported by National Institutes of Health (R01GM112882).

\bibliographystyle{biophysj.bst}
\bibliography{main}

\begin{thebibliography}{46}
\providecommand{\url}[1]{\texttt{#1}}
\providecommand{\urlprefix}{ }

\bibitem[Holliday(1964)]{holliday1964mechanism}
Holliday, R., 1964.
\newblock A mechanism for gene conversion in fungi.
\newblock \emph{Genetical Research} 5:282--304.

\bibitem[Jackson(2002)]{jackson2002sensing}
Jackson, S.~P., 2002.
\newblock Sensing and repairing DNA double-strand breaks.
\newblock \emph{Carcinogenesis} 23:687--696.

\bibitem[Chapman et~al.(2012)Chapman, Taylor, and Boulton]{chapman2012playing}
Chapman, J.~R., M.~R. Taylor, and S.~J. Boulton, 2012.
\newblock Playing the end game: DNA double-strand break repair pathway choice.
\newblock \emph{Molecular cell} 47:497--510.

\bibitem[Li et~al.(2002)Li, Luan, Guo, and Liang]{li2002new}
Li, Q., G.~Luan, Q.~Guo, and J.~Liang, 2002.
\newblock A new class of homogeneous nucleic acid probes based on specific
  displacement hybridization.
\newblock \emph{Nucleic Acids Res.} 30:e5--e5.

\bibitem[Green et~al.(2014)Green, Silver, Collins, and Yin]{green2014toehold}
Green, A.~A., P.~A. Silver, J.~J. Collins, and P.~Yin, 2014.
\newblock Toehold switches: de-novo-designed regulators of gene expression.
\newblock \emph{Cell} 159:925--939.

\bibitem[Jiang et~al.(2015)Jiang, Bhadra, Li, Wu, Milligan, and
  Ellington]{jiang2015robust}
Jiang, Y.~S., S.~Bhadra, B.~Li, Y.~R. Wu, J.~N. Milligan, and A.~D. Ellington,
  2015.
\newblock Robust strand exchange reactions for the sequence-specific, real-time
  detection of nucleic acid amplicons.
\newblock \emph{Anal. Chem.} 87:3314--3320.

\bibitem[Schulman and Doty(2015)]{schulman2015designing}
Schulman, R., and D.~Doty, 2015.
\newblock Designing ordered nucleic acid self-assembly processes.
\newblock \emph{Curr. Opin. Struct. Biol.} 31:57--63.

\bibitem[Zhang et~al.(2012)Zhang, Chen, and Yin]{zhang2012optimizing}
Zhang, D.~Y., S.~X. Chen, and P.~Yin, 2012.
\newblock Optimizing the specificity of nucleic acid hybridization.
\newblock \emph{Nature chemistry} 4:208--214.

\bibitem[Zhang et~al.(2014)Zhang, Hejesen, Kjelstrup, Birkedal, and
  Gothelf]{zhang2014dna}
Zhang, Z., C.~Hejesen, M.~B. Kjelstrup, V.~Birkedal, and K.~V. Gothelf, 2014.
\newblock A DNA-mediated homogeneous binding assay for proteins and small
  molecules.
\newblock \emph{Journal of the American Chemical Society} 136:11115--11120.

\bibitem[Marras et~al.(2015)Marras, Zhou, Su, and
  Castro]{marras2015programmable}
Marras, A.~E., L.~Zhou, H.-J. Su, and C.~E. Castro, 2015.
\newblock Programmable motion of DNA origami mechanisms.
\newblock \emph{PNAS} 112:713--718.

\bibitem[Zhou et~al.(2015)Zhou, Marras, Su, and Castro]{zhou2015direct}
Zhou, L., A.~E. Marras, H.-J. Su, and C.~E. Castro, 2015.
\newblock Direct design of an energy landscape with bistable DNA origami
  mechanisms.
\newblock \emph{Nano Lett.} 15:1815--1821.

\bibitem[Rogers and Manoharan(2014)]{rogers2014programming}
Rogers, W., and V.~Manoharan, 2014.
\newblock Programming colloidal phase transitions with DNA strand displacement.
\newblock \emph{Bull. Am. Phys. Soc.} 59.

\bibitem[Liber et~al.(2015)Liber, Tomov, Tsukanov, Berger, and
  Nir]{liber2015bipedal}
Liber, M., T.~E. Tomov, R.~Tsukanov, Y.~Berger, and E.~Nir, 2015.
\newblock A bipedal DNA motor that travels back and forth between two DNA
  origami tiles.
\newblock \emph{Small} 11:568--575.

\bibitem[Pan et~al.(2015)Pan, Li, Cha, Chen, and Choi]{pan2015recent}
Pan, J., F.~Li, T.-G. Cha, H.~Chen, and J.~H. Choi, 2015.
\newblock Recent progress on DNA based walkers.
\newblock \emph{Curr. Opin. Biotechnol.} 34:56--64.

\bibitem[Landon et~al.(2014)Landon, Lee, Hwang, Mo, Zhang, Neuberger, Meckes,
  Gutierrez, Glinsky, and Lal]{landon2014energetically}
Landon, P.~B., J.~Lee, M.~T. Hwang, A.~H. Mo, C.~Zhang, A.~Neuberger,
  B.~Meckes, J.~J. Gutierrez, G.~Glinsky, and R.~Lal, 2014.
\newblock Energetically Biased DNA Motor Containing a Thermodynamically Stable
  Partial Strand Displacement State.
\newblock \emph{Langmuir} 30:14073--14078.

\bibitem[Tomov et~al.(2013)Tomov, Tsukanov, Liber, Masoud, Plavner, and
  Nir]{tomov2013rational}
Tomov, T.~E., R.~Tsukanov, M.~Liber, R.~Masoud, N.~Plavner, and E.~Nir, 2013.
\newblock Rational design of DNA motors: Fuel optimization through
  single-molecule fluorescence.
\newblock \emph{J. Am. Chem. Soc.} 135:11935--11941.

\bibitem[Masoud et~al.(2012)Masoud, Tsukanov, Tomov, Plavner, Liber, and
  Nir]{masoud2012studying}
Masoud, R., R.~Tsukanov, T.~E. Tomov, N.~Plavner, M.~Liber, and E.~Nir, 2012.
\newblock Studying the structural dynamics of bipedal DNA motors with
  single-molecule fluorescence spectroscopy.
\newblock \emph{ACS Nano} 6:6272--6283.

\bibitem[Li et~al.(2015)Li, Wang, Lu, Chen, Wang, and Shi]{li2015spiking}
Li, X., Z.~Wang, W.~Lu, Z.~Chen, Y.~Wang, and X.~Shi, 2015.
\newblock A Spiking Neural System Based on DNA Strand Displacement.
\newblock \emph{J. Comput. Theor. Nanosci.} 12:298--304.

\bibitem[Wu et~al.(2015)Wu, Wan, Hou, Zhang, Xu, Cui, Wang, Hu, and
  Tan]{wu2015survey}
Wu, C., S.~Wan, W.~Hou, L.~Zhang, J.~Xu, C.~Cui, Y.~Wang, J.~Hu, and W.~Tan,
  2015.
\newblock A survey of advancements in nucleic acid-based logic gates and
  computing for applications in biotechnology and biomedicine.
\newblock \emph{Chem. Commun.} .

\bibitem[Liu et~al.(2014)Liu, Dong, Wu, Fang, Zhou, Shen, Zhou, and
  Hu]{liu2014toehold}
Liu, Y., B.~Dong, Z.~Wu, W.~Fang, G.~Zhou, A.~Shen, X.~Zhou, and J.~Hu, 2014.
\newblock Toehold-mediated DNA logic gates based on host--guest DNA-GNPs.
\newblock \emph{Chem. Commun.} 50:12026--12029.

\bibitem[He et~al.(2014)He, Li, Chen, and Ma]{he2014dna}
He, X., Z.~Li, M.~Chen, and N.~Ma, 2014.
\newblock DNA-Programmed Dynamic Assembly of Quantum Dots for Molecular
  Computation.
\newblock \emph{Angew. Chem. Int. Ed} 126:14675--14678.

\bibitem[Wang et~al.(2014)Wang, Tian, Hou, Ye, and Cui]{wang2014simple}
Wang, Y., G.~Tian, H.~Hou, M.~Ye, and G.~Cui, 2014.
\newblock Simple Logic Computation Based on the DNA Strand Displacement.
\newblock \emph{J. Comput. Theor. Nanosci.} 11:1975--1982.

\bibitem[Qian and Winfree(2011)]{qian2011scaling}
Qian, L., and E.~Winfree, 2011.
\newblock Scaling Up Digital Circuit Computation with DNA Strand Displacement
  Cascades.
\newblock \emph{Science} 332:1196--1201.

\bibitem[Srinivas et~al.(2013)Srinivas, Ouldridge, {\v{S}}ulc, Schaeffer,
  Yurke, Louis, Doye, and Winfree]{srinivas2013biophysics}
Srinivas, N., T.~E. Ouldridge, P.~{\v{S}}ulc, J.~M. Schaeffer, B.~Yurke, A.~A.
  Louis, J.~P. Doye, and E.~Winfree, 2013.
\newblock On the biophysics and kinetics of toehold-mediated DNA strand
  displacement.
\newblock \emph{Nucleic Acids Res.} 41:10641--10658.

\bibitem[Zhang and Winfree(2009)]{zhang2009control}
Zhang, D.~Y., and E.~Winfree, 2009.
\newblock Control of DNA strand displacement kinetics using toehold exchange.
\newblock \emph{J. Am. Chem. Soc.} 131:17303--17314.

\bibitem[Machinek et~al.(2014)Machinek, Ouldridge, Haley, Bath, and
  Turberfield]{machinek2014programmable}
Machinek, R.~R., T.~E. Ouldridge, N.~E. Haley, J.~Bath, and A.~J. Turberfield,
  2014.
\newblock Programmable energy landscapes for kinetic control of DNA strand
  displacement.
\newblock \emph{Nat. Commun.} 5.

\bibitem[Lee et~al.(2014)Lee, von Hippel, and Marcus]{lee2014internally}
Lee, W., P.~H. von Hippel, and A.~H. Marcus, 2014.
\newblock Internally labeled Cy3/Cy5 DNA constructs show greatly enhanced
  photo-stability in single-molecule FRET experiments.
\newblock \emph{Nucleic Acids Res.} gku199.

\bibitem[Le and Kim(2014)]{le2014studying}
Le, T.~T., and H.~D. Kim, 2014.
\newblock Studying DNA looping by single-molecule FRET.
\newblock \emph{JoVE (Journal of Visualized Experiments)} e51667--e51667.

\bibitem[Aitken et~al.(2008)Aitken, Marshall, and Puglisi]{aitken2008oxygen}
Aitken, C.~E., R.~A. Marshall, and J.~D. Puglisi, 2008.
\newblock An oxygen scavenging system for improvement of dye stability in
  single-molecule fluorescence experiments.
\newblock \emph{Biophys. J.} 94:1826--1835.

\bibitem[Bel et~al.(2010)Bel, Munsky, and Nemenman]{bel2010simplicity}
Bel, G., B.~Munsky, and I.~Nemenman, 2010.
\newblock The simplicity of completion time distributions for common complex
  biochemical processes.
\newblock \emph{Physical biology} 7:016003.

\bibitem[Anshelevich et~al.(1984)Anshelevich, Vologodskii, Lukashin, and
  Frank-Kamenetskii]{anshelevich1984slow}
Anshelevich, V., A.~Vologodskii, A.~Lukashin, and M.~Frank-Kamenetskii, 1984.
\newblock Slow relaxational processes in the melting of linear biopolymers: a
  theory and its application to nucleic acids.
\newblock \emph{Biopolymers} 23:39--58.

\bibitem[Kim(1958)]{kim1958mean}
Kim, S.~K., 1958.
\newblock Mean first passage time for a random walker and its application to
  chemical kinetics.
\newblock \emph{J. Chem. Phys.} 28:1057--1067.

\bibitem[Murthy and Kehr(1989)]{murthy1989mean}
Murthy, K. P.~N., and K.~W. Kehr, 1989.
\newblock Mean first-passage time of random walks on a random lattice.
\newblock \emph{Phys. Rev. A} 40:2082--2087.

\bibitem[Pury and C{\'a}ceres(2003)]{pury2003mean}
Pury, P.~A., and M.~O. C{\'a}ceres, 2003.
\newblock Mean first-passage and residence times of random walks on asymmetric
  disordered chains.
\newblock \emph{Journal of Physics A: Mathematical and General} 36:2695.

\bibitem[Lakatos et~al.(2004)Lakatos, Chou, Bergersen, and
  Patey]{lakatos2004first}
Lakatos, G., T.~Chou, B.~Bergersen, and G.~Patey, 2004.
\newblock First passage times of pulling-assisted DNA unzipping.
\newblock \emph{arXiv preprint cond-mat/0406006} .

\bibitem[Gao et~al.(2006)Gao, Wolf, and Georgiadis]{gao2006secondary}
Gao, Y., L.~K. Wolf, and R.~M. Georgiadis, 2006.
\newblock Secondary structure effects on DNA hybridization kinetics: a solution
  versus surface comparison.
\newblock \emph{Nucleic Acids Res.} 34:3370--3377.

\bibitem[Cisse et~al.(2012)Cisse, Kim, and Ha]{cisse2012rule}
Cisse, I.~I., H.~Kim, and T.~Ha, 2012.
\newblock A rule of seven in Watson-Crick base-pairing of mismatched sequences.
\newblock \emph{Nat. Struct. Mol. Biol.} 19:623--627.

\bibitem[Cocco et~al.(2001)Cocco, Monasson, and Marko]{cocco2001force}
Cocco, S., R.~Monasson, and J.~F. Marko, 2001.
\newblock Force and kinetic barriers to unzipping of the DNA double helix.
\newblock \emph{PNAS} 98:8608--8613.

\bibitem[Van~Kampen(1992)]{van1992stochastic}
Van~Kampen, N.~G., 1992.
\newblock Stochastic processes in physics and chemistry, volume~1.
\newblock Elsevier.

\bibitem[Floyd et~al.(2010)Floyd, Harrison, and Van~Oijen]{floyd2010analysis}
Floyd, D.~L., S.~C. Harrison, and A.~M. Van~Oijen, 2010.
\newblock Analysis of kinetic intermediates in single-particle dwell-time
  distributions.
\newblock \emph{Biophysical journal} 99:360--366.

\bibitem[Betterton and J\"ulicher(2005)]{betterton2005opening}
Betterton, M.~D., and F.~J\"ulicher, 2005.
\newblock Opening of nucleic-acid double strands by helicases: Active versus
  passive opening.
\newblock \emph{Phys. Rev. E} 71:011904.

\bibitem[Panyutin and Hsieh(1994)]{panyutin1994kinetics}
Panyutin, I.~G., and P.~Hsieh, 1994.
\newblock The kinetics of spontaneous DNA branch migration.
\newblock \emph{PNAS} 91:2021--2025.

\bibitem[Waters and Kim(2015)]{waters2015calculation}
Waters, J.~T., and H.~D. Kim, 2015.
\newblock Calculation of a fluctuating entropic force by phase space sampling.
\newblock \emph{Physical Review E} 92:013308.

\bibitem[Paramanathan et~al.(2014)Paramanathan, Reeves, Friedman, Kondev, and
  Gelles]{paramanathan2014general}
Paramanathan, T., D.~Reeves, L.~J. Friedman, J.~Kondev, and J.~Gelles, 2014.
\newblock A general mechanism for competitor-induced dissociation of molecular
  complexes.
\newblock \emph{Nat. Commun.} 5.

\bibitem[Zuker(2003)]{zuker2003mfold}
Zuker, M., 2003.
\newblock Mfold web server for nucleic acid folding and hybridization
  prediction.
\newblock \emph{Nucleic Acids Res.} 31:3406--3415.

\bibitem[Mentes et~al.(2015)Mentes, Florescu, Brunk, Wereszczynski, Joyeux, and
  Andricioaei]{mentes2015free}
Mentes, A., A.~M. Florescu, E.~Brunk, J.~Wereszczynski, M.~Joyeux, and
  I.~Andricioaei, 2015.
\newblock Free-Energy Landscape and Characteristic Forces for the Initiation of
  DNA Unzipping.
\newblock \emph{Biophys. J.} 108:1727--1738.

\end{thebibliography}

\clearpage
\newcommand{\beginsupplement}{%
        \setcounter{table}{0}
        \renewcommand{\thetable}{S\arabic{table}}%
        \setcounter{figure}{0}
        \renewcommand{\thefigure}{S\arabic{figure}}%
        \setcounter{equation}{0}
        \renewcommand{\theequation}{S\arabic{equation}}%
        \renewcommand{\figurename}{Supplementary Figure}
        \renewcommand{\tablename}{Supplementary Table}
        \setcounter{page}{1}
}


\beginsupplement
\onecolumngrid
\section*{Supporting Material}

\subsection*{Oligonucleotide sequences used in this study}

\begin{table}[h!]
\begin{center}
\begin{tabular}{ l c }
Substrate  & 5'-CA/iCy5/ACCAAAATTGGGACAACACCAGTG/3BioTEG/-3' \\
Substrate* & 5'-CA/iCy5/ATTAAAATTCCGACAACACCAGGT/3BioTEG/-3' \\
\end{tabular}
\end{center}
\caption{Substrate strand. The substrate strand sequence is complementary to a region of messenger RNA encoding YFP. We internally labelled the strand with Cy5 to increase photostability\cite{lee2014internally} and implemented a biotin linker for surface immobilization. The substrate* strand was derived from the substrate and altered to remove secondary structure.}
\label{table:immob-strands}
\end{table}

\begin{table}[h!]
\begin{center}
\begin{tabular}{ l c }
 Match        & 5'-GTCCCAATTTTGGT/BHQ3/-3' \\
 Match*       & 5'-GTCGGAATTTTAAT/BHQ3/-3' \\
 Mismatch  1  & 5'-\underline{A}TCCCAATTTTGGT/BHQ3/-3' \\
 Mismatch  2  & 5'-G\underline{C}CCCAATTTTGGT/BHQ3/-3' \\
 Mismatch  3  & 5'-GT\underline{T}CCAATTTTGGT/BHQ3/-3' \\
 Mismatch  4  & 5'-GTC\underline{T}CAATTTTGGT/BHQ3/-3' \\
 Mismatch  5  & 5'-GTCC\underline{T}AATTTTGGT/BHQ3/-3' \\
 Mismatch  6  & 5'-GTCCC\underline{G}ATTTTGGT/BHQ3/-3' \\
 Mismatch  7  & 5'-GTCCCA\underline{G}TTTTGGT/BHQ3/-3' \\
 Mismatch  8  & 5'-GTCCCAA\underline{C}TTTGGT/BHQ3/-3' \\
 Mismatch  9  & 5'-GTCCCAAT\underline{C}TTGGT/BHQ3/-3' \\
 Mismatch 10  & 5'-GTCCCAATT\underline{C}TGGT/BHQ3/-3' \\
 Mismatch 11  & 5'-GTCCCAATTT\underline{C}GGT/BHQ3/-3' \\
 Mismatch 12  & 5'-GTCCCAATTTT\underline{A}GT/BHQ3/-3' \\
 Mismatch 13  & 5'-GTCCCAATTTTG\underline{A}T/BHQ3/-3' \\
 Mismatch 14  & 5'-GTCCCAATTTTGG\underline{C}/BHQ3/-3' \\
\end{tabular}
\end{center}
\caption{Incumbent strands. The incumbent strand was labelled with a dark quencher with an absorption spectrum that well overlaps the emission of Cy5. The underlined letter represents the single mismatch. The match* strand was designed to remove secondary structure.}
\label{table:inc-strands}
\end{table}

\begin{table}[h!]
\begin{center}
\begin{tabular}{ l c }
 Match        & 5'-CACTGGTGTTGTCCCAATTTTGGT-3' \\
 Match*       & 5'-ACCTGGTGTTGTCGGAATTTTAAT-3' \\
 Mismatch  1  & 5'-CACTGGTGTT\underline{A}TCCCAATTTTGGT-3' \\
 Mismatch  1* & 5'-ACCTGGTGTT\underline{A}TCGGAATTTTAAT-3' \\ 
 Mismatch  2  & 5'-CACTGGTGTTG\underline{C}CCCAATTTTGGT-3' \\
 Mismatch  3  & 5'-CACTGGTGTTGT\underline{T}CCAATTTTGGT-3' \\
 Mismatch  4  & 5'-CACTGGTGTTGTC\underline{T}CAATTTTGGT-3' \\
 Mismatch  5  & 5'-CACTGGTGTTGTCC\underline{T}AATTTTGGT-3' \\
 Mismatch  6  & 5'-CACTGGTGTTGTCCC\underline{G}ATTTTGGT-3' \\
 Mismatch  7  & 5'-CACTGGTGTTGTCCCA\underline{G}TTTTGGT-3' \\
 Mismatch  7* & 5'-ACCTGGTGTTGTCGGA\underline{G}TTTTAAT-3' \\
 Mismatch  8  & 5'-CACTGGTGTTGTCCCAA\underline{C}TTTGGT-3' \\
 Mismatch  9  & 5'-CACTGGTGTTGTCCCAAT\underline{C}TTGGT-3' \\
 Mismatch 10  & 5'-CACTGGTGTTGTCCCAATT\underline{C}TGGT-3' \\
 Mismatch 11  & 5'-CACTGGTGTTGTCCCAATTT\underline{C}GGT-3' \\
 Mismatch 12  & 5'-CACTGGTGTTGTCCCAATTTT\underline{A}GT-3' \\
 Mismatch 13  & 5'-CACTGGTGTTGTCCCAATTTTG\underline{A}T-3' \\
 Mismatch 14  & 5'-CACTGGTGTTGTCCCAATTTTGG\underline{C}-3' \\
\end{tabular}
\end{center}
\caption{Invader strands. The underlined letter represents the single mismatch. The strands marked with an asterisk were designed by removing secondary structure from their corresponding invader strands.}
\label{table:inv-strands}
\end{table}

\subsection*{Branch migration as random walk}

We put forth a model for strand displacement based on the mean first passage time of a 1D random walk. We begin by assuming that the rate of breaking individual base pairs is much slower than the reverse rate of formation. By this assumption, incumbent strand unzipping and invader strand zipping is almost coincidental, and intermediates states can be specified with one state variable $i$, which is equal to the number of displaced base pairs. $i=0$ is the initial state before invasion, and $i=n$ corresponds to complete displacement. We now define a 1D lattice with $n+1$ sites. Motion at i-th lattice site is performed in single steps at forward and reverse rates, $f_i$ and $r_i$ respectively. This model is equivalent to a random walk with a perfectly reflecting boundary on the left ($i=0$), and perfectly absorbing boundary on the right ($i=n$). 

The mean first passage time from $i=0$ to $i=n$ is given by\cite{kim1958mean,murthy1989mean}
\begin{equation} \label{eq:1dmatch}
\tau = \sum_{i=0}^{n-1} \frac{1}{p_if_i} ,
\end{equation}
where $p_i$ is the steady state probability at site $i$ in a partial lattice between $0$ and $i$. Therefore, the inverse of each term, $p_if_i$, can be interpreted as the effective rate of reaching $i+1$ from an unspecified previous position. $p_i$ can be expressed with a ratio of forward and reverse rates between two adjacent sites ($\alpha_i=f_{i-1}/r_i$) as
\begin{equation}
p_i=\frac{\alpha_i\alpha_{i-1}...1}{1+\alpha_1+\alpha_2\alpha_1+...+\alpha_i\alpha_{i-1}...1}
\label{seq:1dmfpt}
\end{equation}

Without sequence dependence, branch migration over a matched base pair must be identical in either direction and, therefore, $f_{i-1} = r_i$ or $\alpha_i=1$. In comparison, $\alpha_i \gg 1$ for the case of a mismatch on the incumbent and $\alpha_i \ll 1$ for the case of a mismatch in the invader. We denote this mismatch-dependent fold-change in $\alpha$ as $a$, which must be larger than one for an incumbent mismatch and smaller than one for an invader mismatch. We also introduce variation in the forward rate for the first base pair to be displaced with another ratio ($f/b$). It is thought to be smaller due to slow initiation ($b>1$)\cite{mentes2015free}. Using these ratios, the MFPT's with an invader mismatch ($\tau_v$) and an incumbent mismatch ($\tau_c$) at position $j$ are given by
\begin{subequations}
\begin{equation}
\tau_v(j)=\frac{1}{f}\left[-\left(\frac{1}{a}-1\right)j^2+n\left(\frac{1}{a}-1\right)j+(b-1)\left(-\left(\frac{1}{a}-1\right)j+\frac{n}{a}\right)+\frac{n(n+1)}{2}\right] ,
\end{equation}
\begin{equation}
\tau_c(j)=\frac{1}{f}\left[(1-a)j^2-(1-a)(n+1)j+(b-1)((1-a)(j-1)+an)+\frac{n(n+1)}{2}\right] ,
\end{equation}
\label{eq:1dmodel}
\end{subequations}
respectively. The equations are cast in a form to reveal the dependence of MFPT on mismatch position $j$. Without slower opening of the first base pair ($b=1$), MFPT for the invader mismatch is concave down with a center at $n/2$, and MFPT for the incumbent mismatch is concave up with a center at $(n+1)/2$. Slow opening of the first base pair ($b>1$) shifts the center towards lower values. As expected, when $a=1$ and $b=1$, both MFPT's approach $n^2/2f$.

This 1D model predicts MFPT to be a quadratic function of mismatch position  with the slowest displacement near the center position (Fig.~\ref{sfig:1dmodel}(a)), which is not consistent with the overall monotonic change we observed with an invader mismatch. Slow initiation of branch migration ($b>1$) could render the prediction more monotonic (Fig.~\ref{sfig:1dmodel}(b)), but it requires an unreasonably large $b$.    


\subsection*{Dissociation kinetics of the invader strand}
In effort to instill confidence in our model, we performed a separate experiment involving biotinylated invader strands, Cy3 labelled substrate strands, and Cy5 labelled incumbent strands. The invader strands were immobilized on the surface, and preformed substrate-incumbent duplexes were pumped in and allowed to react. Invader strands were designed to have the same complementary toehold (10nt) adjoined to a tail composed of 14 thymidines to prevent successful strand displacement. Interactions between duplexes and invader strands were recognized as a high FRET state. Lifetimes of high FRET states were recorded and interpreted as dissociation times. Large numbers of these lifetimes were recorded to construct a single exponential probability distribution whose decay rate was determined to be \SI{\sim0.03}{\per\second}. 

\begin{figure}[!th]
\begin{minipage}[c][\textheight]{\textwidth}
\includegraphics[width=17cm]{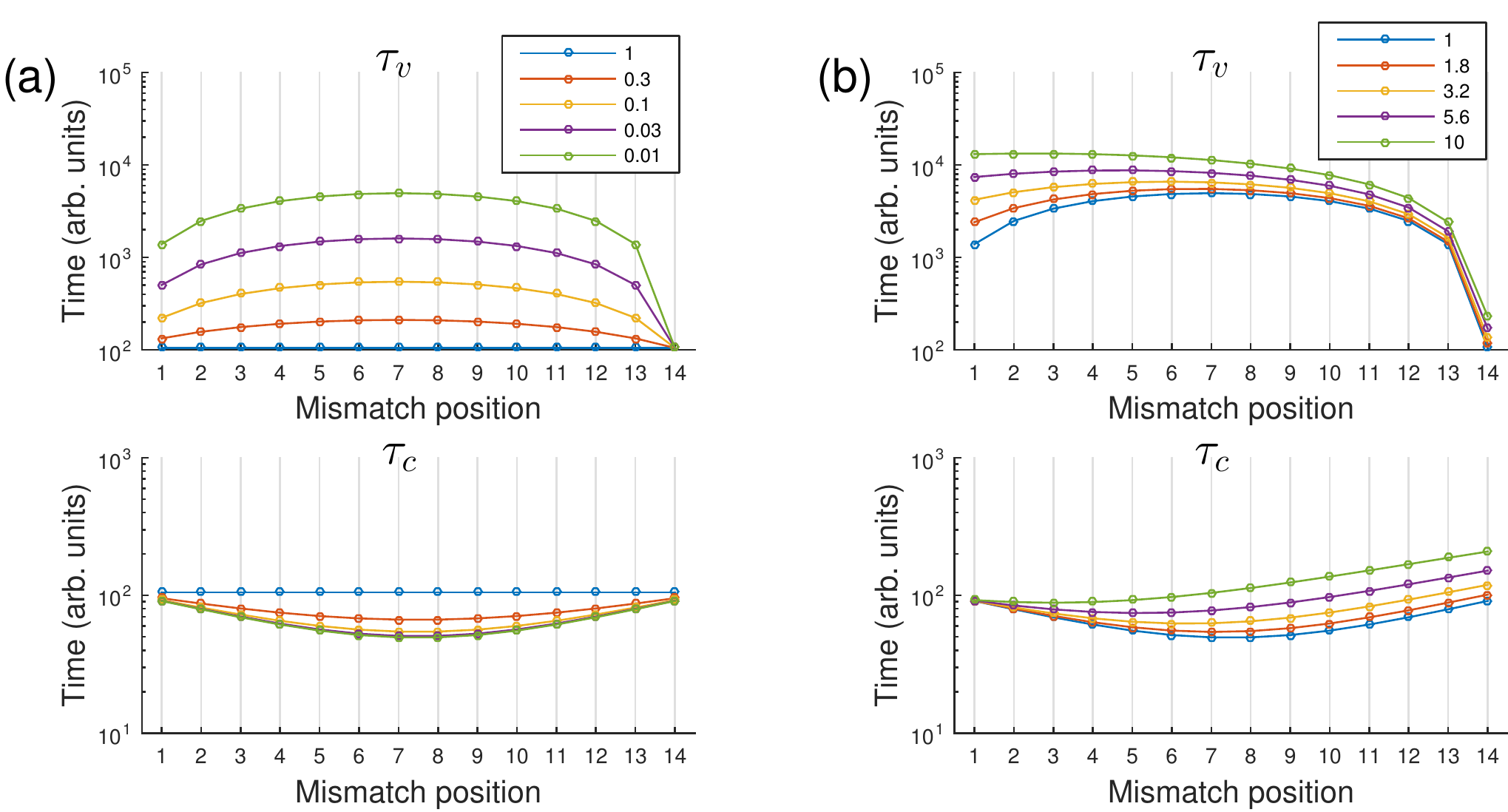}
\caption{Dissociation via branch migration. MFPT's predicted by the 1D lattice model are plotted using Eq.~\ref{eq:1dmodel}. (a) MFPT's with varying $a$. $a$ characterizes the effect of a mismatch on the forward branch migration rate. A mismatch in the invading strand lengthens MFPT ($\tau_v$), while a mismatch in the incumbent strand shortens it ($\tau_v$). $b$ is fixed to $1$. (b) MFPT's with varying $b$. $b$ represents how slow the first migration step is compared to the rest. As $b$ becomes larger, the center of the curves shifts towards the left, and both $\tau_v$ and $\tau_c$ become more monotonic as a function of mismatch position. $a$ is fixed to $0.01$.}
\label{sfig:1dmodel}
\end{minipage}
\end{figure}

\begin{figure}[!th]
\begin{minipage}[c][\textheight]{\textwidth}
\includegraphics[width=17cm]{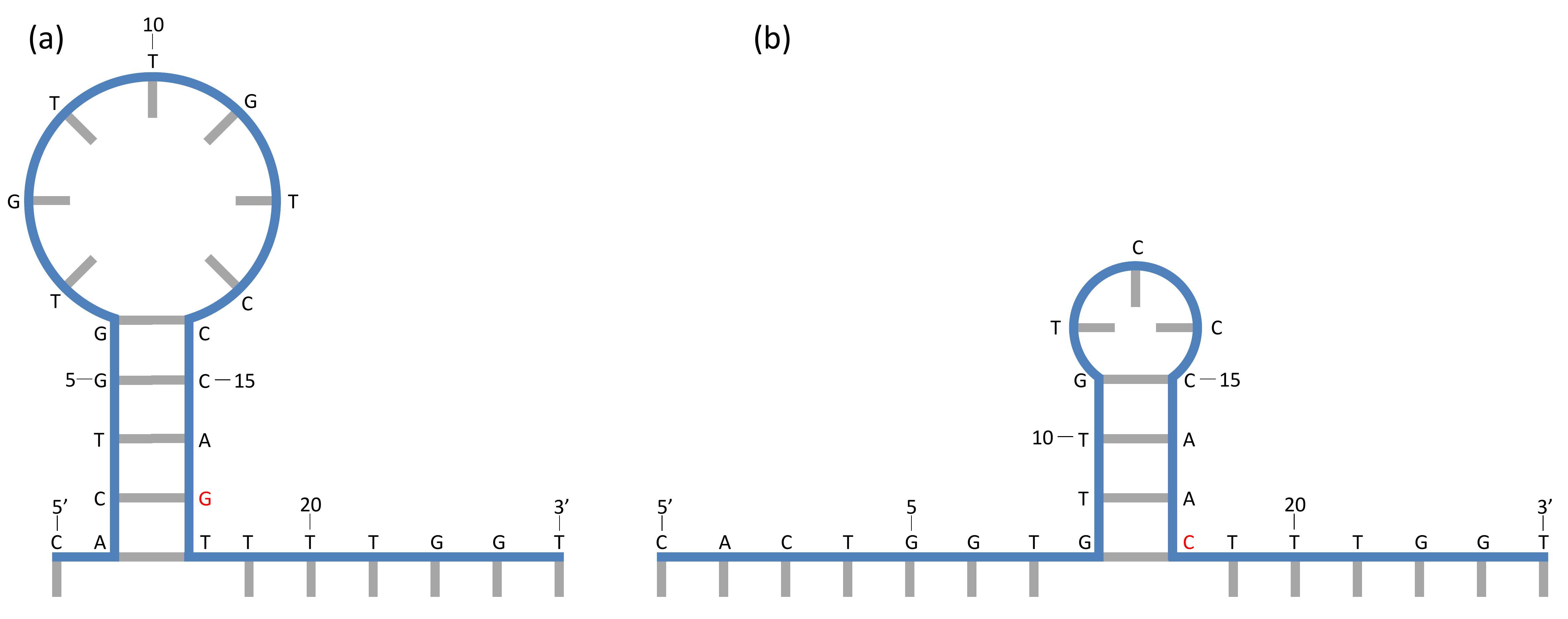}
\caption{Putative secondary structures of invader mismatch strands. (a) Mismatch position 7. This is the only conformation predicted by mfold\cite{zuker2003mfold} for this structure. The hairpin mostly obstructs the toehold. It is predicted to have a lower free energy than the active form by $\Delta G=\SI{-3.61}{kcal \per mol}$. (b) Mismatch position 8.  This is the predicted conformation by mfold. The toehold is partially obstructed due to the hairpin. It is predicted to have a lower free energy from the active form by $\Delta G=\SI{-2.61}{kcal \per mol}$.}
\label{sfig:secondary-struct}
\end{minipage}
\end{figure}

\begin{figure}[!th]
\begin{minipage}[c][\textheight]{\textwidth}
\includegraphics[width=10cm]{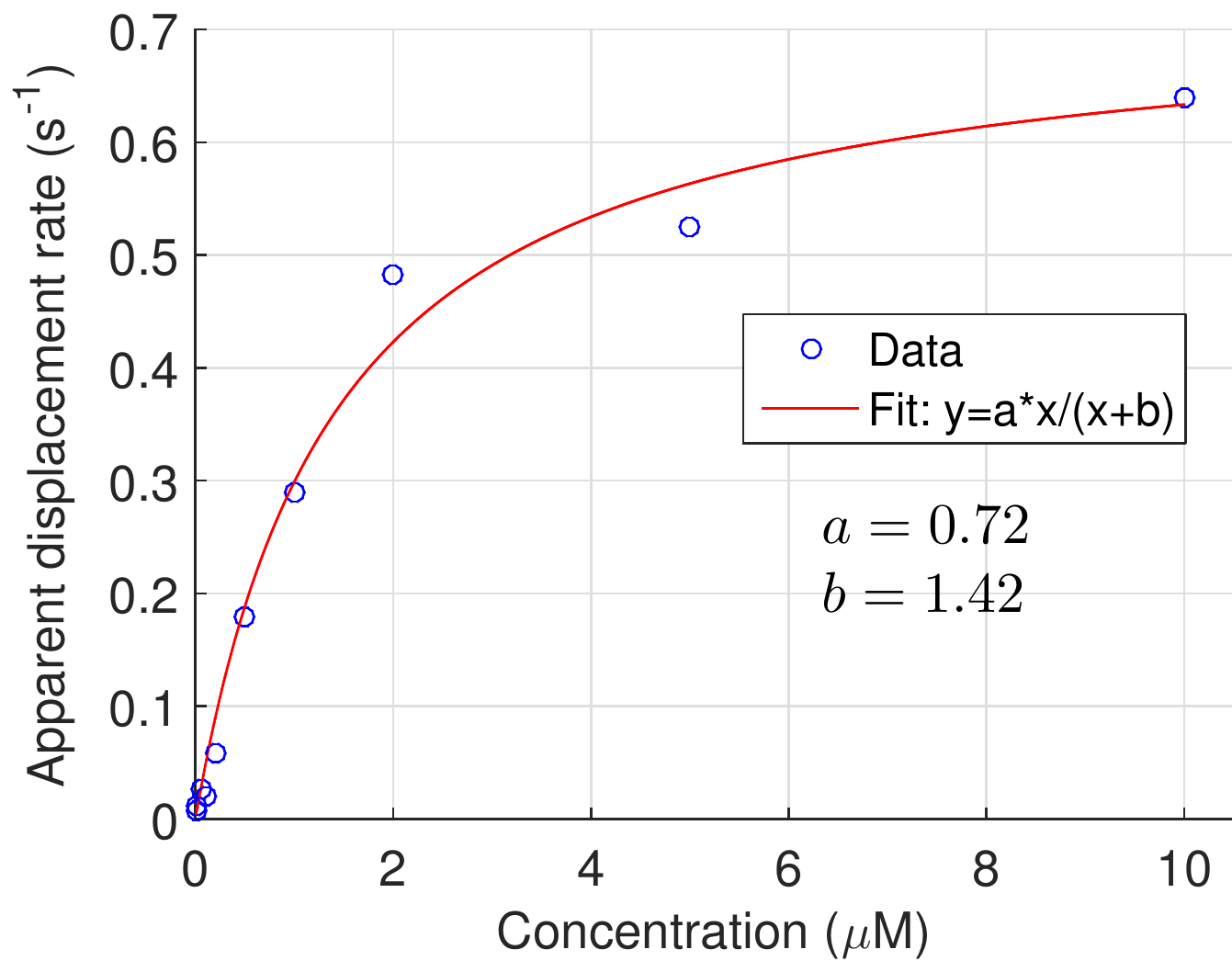}
\caption{Displacement rate vs. invader concentration. The displacement rate ($y$) was measured as a function of the concentration of the invader strand ($x$). The displacement reaction can be modeled by reversible binding ($k_+$) and unbinding ($k_-$) steps followed by a unimolecular displacement ($r$) step. The unbinding rate of the toehold-bound invader was directly measured to be $\SI{1/30}{\per \s}$, which is much slower than $r$. In this case, the apparent displacement rate is given by $rx/(x+r/k_+)$. We fit the measured data points (blue hollow circles) using the expression $y=ax/(x+b)$ with two fitting parameters. From $a$ and $b$, we determine the unimolecular displacement rate ($r$) and the binding rate ($k_+$) to be $\SI{0.72}{\per \s}$ and $\SI{0.5}{\per \micro \molar \per \s}$, respectively.}
\label{sfig:binding}
\end{minipage}
\end{figure}
\end{document}